# COGEDAP: A COmprehensive GEnomic Data Analysis Platform

Bayram Cevdet Akdeniz1,2, Oleksandr Frei1,2, Espen Hagen2, Tahir Tekin Filiz2, Sandeep Karthikeyan2, Joëlle Pasman3, Andreas Jangmo4, Jacob Bergstedt3, John R. Shorter5,6, Richard Zetterberg5,6, Joeri Meijsen5,6, Ida Elken Sønderby2,7, Alfonso Buil5,6, Martin Tesli4, Yi Lu3, Patrick Sullivan3, Ole Andreassen2, Eivind Hovig1,8

Affiliations:
1. Center for Bioinformatics, Department of Informatics, University of Oslo, Oslo, Norway
2. Norwegian Centre for Mental Disorders Research (NORMENT), Division of Mental Health and Addiction, Oslo University Hospital and University of Oslo, Oslo, Norway
3. Department of Medical Epidemiology and Biostatistics, Karolinska Institutet
4. Department of Mental Disorders, Norwegian Institute of Public Health, Oslo, Norway
5. Institute of Biological Psychiatry, Mental Health Center Sct. Hans, Mental Health
Services Copenhagen, Roskilde, Denmark.
6. The Lundbeck Foundation Initiative for Integrative Psychiatric Research (iPSYCH),
Copenhagen, Denmark.
7. Department of Medical Genetics, Oslo University Hospital, Oslo, Norway
8. Department of Tumor Biology, Institute for Cancer Research, Oslo University Hospital, Oslo, Norway

## ABSTRACT

Non-sharable sensitive data collection and analysis in large-scale consortia for genomic research is complicated. Time-consuming issues in installing software arise due to different operating systems, software dependencies and running the software. Therefore, easier, more standardized, automated protocols and platforms can be a solution to overcome these issues. We have developed one such solution for genomic data analysis using software container technologies. The platform, "COGEDAP", consists of different software tools placed into Singularity containers with corresponding pipelines and instructions on how to perform genome-wide association studies (GWAS) and other genomic data analysis via corresponding tools. Using a provided helper script written in Python, users can obtain auto-generated scripts to conduct the desired analysis both on high-performance computing (HPC) systems and on personal computers. The analyses can be done by running these auto-generated scripts with the software containers. The helper script also performs minor re-formatting of the input/output data, so that the end user can work with a unified file format regardless of which genetic software is used for the analysis. COGEDAP is actively being used by users from different countries/projects to conduct their genomic data analyses. Thanks to this platform, users can easily run GWAS and other genomic analyses without spending much effort on software installation, data formats, and other technical requirements.

I. INTRODUCTION

Genome-wide association studies (GWAS) represent a widely used approach to determine the relationship between a trait of interest and genomic data [20]. In the last decade, many GWAS approaches, and corresponding tools, have been developed for different applications. Some of the most widely used GWAS tools are PLINK [6-7], GCTA [8], BOLT-LMM [9], SAIGE [10], and REGENIE [11]. In addition to the GWAS analysis tools, supplementary tools/packages for pre-processing or post-processing data are often required. An additional issue is the spending time to understand the correct parameters and options needed to operate these computational tools. Despite some common terminology and syntax, since all these tools have been created by different developers, their required inputs in terms of data formats and flags are often different. Furthermore, installing all these tools natively on a machine can sometimes be challenging, since users may face with some issues related to the operating systems and software dependencies. Another challenge in sensitive data analysis is that of data sharing, due to the General Data Protection Regulation (GDPR). This requires the containment of sensitive genomic data into secure systems with controlled and limited internet access. Because of these restrictions, it may be challenging to have the required packages/tools available on such secure systems to conduct the desired analysis. It becomes even more challenging if several sites are willing to do a joint analysis with their genomic data. Since genomic data often cannot be shared among the sites, one straightforward solution is to apply the exact same methodology and computation on the data, and then combine and/or compare the results. Therefore, having a data analysis platform that can work on different systems enables standard analysis on different systems at different sites.

One application of this approach has been developed to identify causes of comorbidity between mental disorders and cardiovascular disease [21]. Many institutions from different countries have collaborated with their non-shareable datasets and different infrastructures for the data analysis aspects of the questions to be addressed. We developed an analysis platform for distributed genomic analysis which is demonstrated to function at all sites. Prior to analysis, data harmonization was required across these sites to combine cohorts and replicate results, thereby fully benefitting from the large number of cohorts existing in Estonia, Finland, Scotland, and the Nordic countries.

The use of software container technologies greatly facilitated these goals, in a scalable and sustainable fashion for other similar GWAS datasets and research questions. We here refer to this data platform as COmprehensive GEnomic Data Analysis Platform (COGEDAP) and it is now actively being used for a number of genomic analysis projects.

We are aware of already existing pipelines that are aimed to solve some of the issues listed above. In [1], an automated GWAS pipeline called nf-GWAS, implemented in Nextflow and distributed with Docker containers, was developed using R-based tools including SNPRelate/GENESIS/GMMAT and ANNOVAR. Another Nextflow-based pipeline that uses Regenie for GWAS has been developed in [2]. A GWAS pipeline called GWASpi, which is a JAVA-based platform to perform GWAS analysis using PLINK has been developed both as a web-based and command-line based platform [3]. In [4], a GWAS pipeline has been developed using the tools TASSEL and GAPIT. The RICOPILI pipeline proposed in [5] can be considered as one of the most comprehensive pipelines among the existing pipelines, including Quality Control (QC), GWAS analysis, and imputation.

One of the main assets of our data platform compared to the previous works is its versatility both in terms of analysis and available tools. For instance, our data platform enables conducting different genomic data analysis, like QC, GWAS, imputation, Polygenic Risk Scores (PRS) in a manner somewhat similar to that of Ricopili but our platform offers greater versatility in the different tools for a given analysis. For instance, COGEDAP has presently been made capable of performing GWAS with five different tools, including PLINK, GCTA, BOLT-LMM, SAIGE, and REGENIE. The more comprehensive nature of COGEDAP offers a more flexible approach compared to other existing pipelines in the literature. Furthermore, our data platform is both personal computer (PC)- and HPC- friendly, enabling auto-generated scripts to run the analysis on both system types.

In the following sections, we first define the methodology regarding the development of the COGEDAP data platform, followed by a presentation of the installation and workflow of the COGEDAP. Subsequently, we present a use case related to performing a GWAS with different tools available within COGEDAP, and finally, we discuss future development directions.

## II. METHODOLOGY & TECHNICAL REQUIREMENTS

Firstly, in order to achieve such data platform, the main requirements (or specification) of the platform for data harmonization and analysis were determined as:

1. Compliance with GDPR regulations and with secure handling of sensitive genomic data,
2. Portable, easily integrated with each partner's infrastructure, which often involves servers with limited internet access
3. Standardized data formats and software, enabling the methods reproducibility without version differences
4. Automated, easy to run without spending time and effort

Then, to find a solution for enabling cross-border analyses with desired properties, we examined a pre-existing prototypes in Tryggve2 project [22]. Based on this survey, we concluded that the use of software container technologies would be the best technical solution for the platform requirements defined above. GWAS and other tools embedded into the containers and accompanied by clear usage guidelines are decided to be a part of our solution for distributed data analysis, as depicted in Figure 1. Users at any participating site can download the software containers from the central Github repository[1], along with the required instructions to execute the corresponding analysis. Further, the containers were made to function with high-performance computing (HPC) systems with limited internet access to operate on the sensitive data hosted at each site. The proposed method can be regarded as secure in GDPR sense, since sensitive individual-level data never leave each partner's storage system. Also, it is portable, since each site can easily upload and use containers directly. It is standardized, in the sense that each site will be using the same embedded software tools with identical version numbers and sophisticated pipelines. As a practical demonstration of the COGEDAP platform, we will throughout this manuscript showcase the (semi)automated generation of analysis scripts and corresponding compute jobs for different GWAS applications. The amount of user input is minimalized in contrast to standard setups.

---

[1] https://github.com/comorment/containers/tree/main/singularity

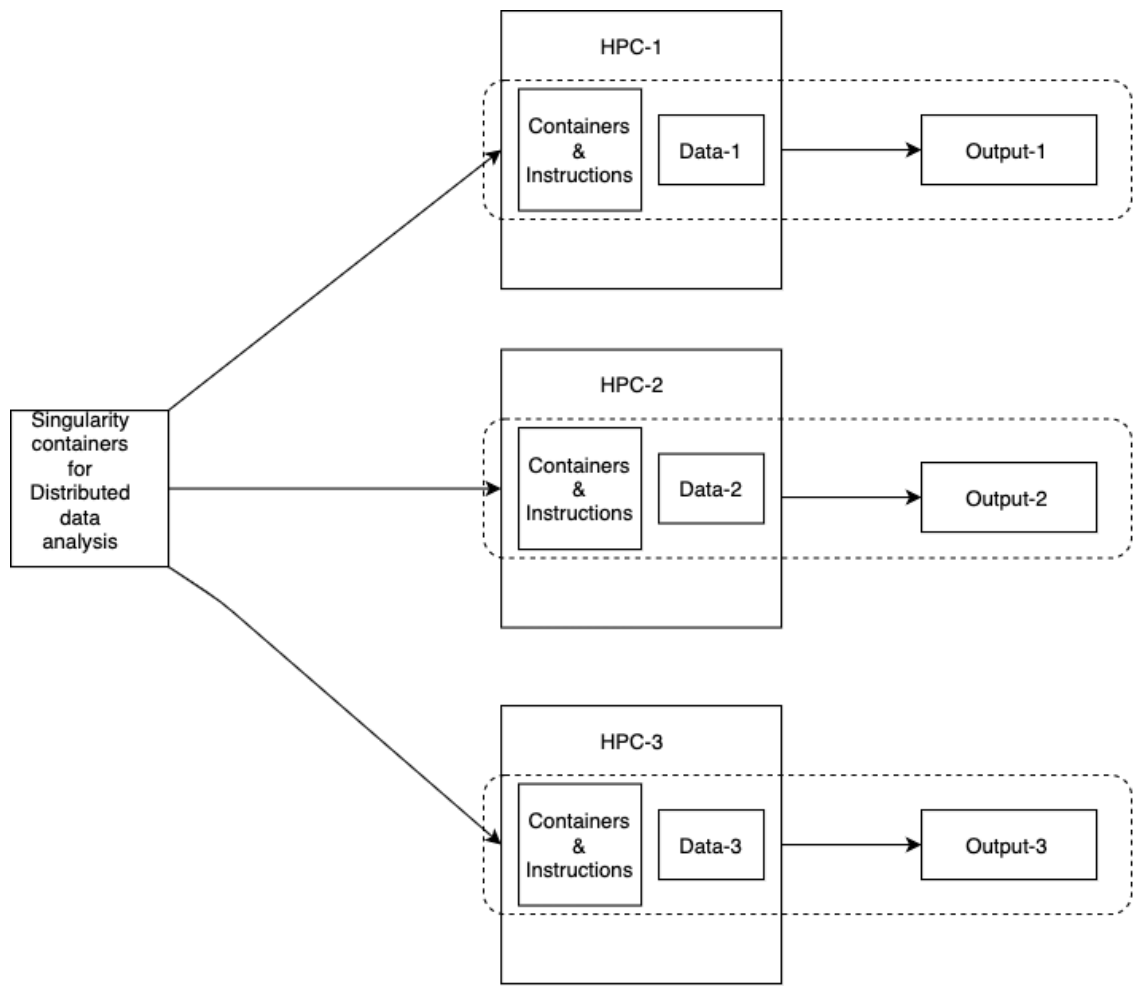

Figure 1. Diagram for distributed data analysis using software containers.

We chose to use Singularity (https://docs.sylabs.io) as our container-based solution. Docker is the most popular container platform at the time of writing, having extensive documentation and a large user community. On the other hand, Singularity containers are more suitable for HPC use, as for instance, they do not require root privileges which is an important trait while using HPC systems. Furthermore, Singularity containers can easily be built via Docker containers. This means we can benefit from the Docker container documentation and large community for developing our Singularity containers.

In order to verify that this solution was applicable at all sites, we determined the technical requirements to run our designated analysis and prepared a demo for distributed analysis via Singularity containers with instructions on how to run it on HPC systems without internet access. The demo container includes the PLINK software with some instructions to run it on HPC, and in order to run it on an HPC system, the following system properties are required:

1- A Linux or Unix-like operating system on the HPC system

2- The installation of Singularity (https://docs.sylabs.io), being version 3 or higher, on the HPC system.

3- A SLURM (https://slurm.schedmd.com/documentation.html) or Sun Grid Engine (SGE, https://docs.oracle.com/cd/E19279-01/820-3257-12/n1ge.html) HPC scheduler

Note that for running the platform on a local machine, only the first two requirements are necessary.

## III. APPLICATION OF THE COGEDAP PLATFORM

The descriptions of installation of the COGEDAP, available tools, basic workflow, and a use case that shows how to conduct data analysis with a demo GWAS analysis example are provided below. It is important to note that the links related to COGEDAP provided here can also be referenced from [25]. We are going to mention some sub-directories in this repository to define some aspects of the platform below. For instance, if we refer to docs/hello.md, this is the "hello.md" file located under the "docs" folder of our repository (https://github.com/comorment/containers). At the time of writing, this manuscript corresponds to the release of version 1.1 [25].

### a) Installation

For installing the data platform, we recommend cloning the entire code repository (https://github.com/comorment/containers.git) using Git (https://git-scm.com). To do this Git Large File System (LFS) extension beforehand (https://git-lfs.github.com/) should be enabled. The details of the installation can be found in our GitHub repository (https://github.com/comorment/containers#getting-started).

Once a clone of this repository has been made available on the system, a provided suite of unit tests has been implemented using the Pytest framework (https://pytest.org) for running basic checks that all supplied software is present within each respective container. The installation of Pytest and test runs can be performed in a local Python environment on the host by issuing the following:

```
$ cd <path/to/containers>  # change directory to repository root
$     pip     install     pytest     #     installs     Pytest
$ py.test -v tests # runs test suite
```

Then, there is a demo example in the repository placed on the docs/hello.md, and it can be used to test that the data platform works on the system.

### b) Available tools

One of the important properties of COGEDAP is its versatility, achieved by enabling different tools to run easily. A complete, up-to-date list of tools available in our data platform is available in the CHANGELOG.md file. Table 1 lists some of the most widely used tools.

| ANALYSIS | TOOLS |
|---|---|
| GWAS analysis | Plink [6], Plink2 [7], Regenie [11], Saige [10], Bolt-LMM [9], GCTA [8] |
| Quality Control (QC) of Genotyping Data | pipeline for complete QC of genotyping data (Moba QC) pipeline [14] |
| Imputation | MoBa Imputation pipeline [14] |
| Post-GWAS analysis | LD SCore[15], qq.py manhattan.py [13] |
| Polygenic Risk Score calculation | PRSice 2 [17], SBAYESR [18], LDPRED2 [19] |
| Meta Analysis | metal [16] |
| Other tools | pleiofdr, mixer, mostest [12] |

Table 1. List of analysis and corresponding software tools.

### c) Workflow

Each workflow to run genomic analysis with our data platform consists of some basic steps, including organizing the data into a standardized format, running the Python-implemented helper script gwas.py to generate scripts for the corresponding data analysis, running the auto-generated scripts for the analysis and further post-processing of the analysis results. These steps for GWAS analysis can be detailed in Workflow 1. Note that similar workflows for different kinds of analyses are also available on the COGEDAP website.

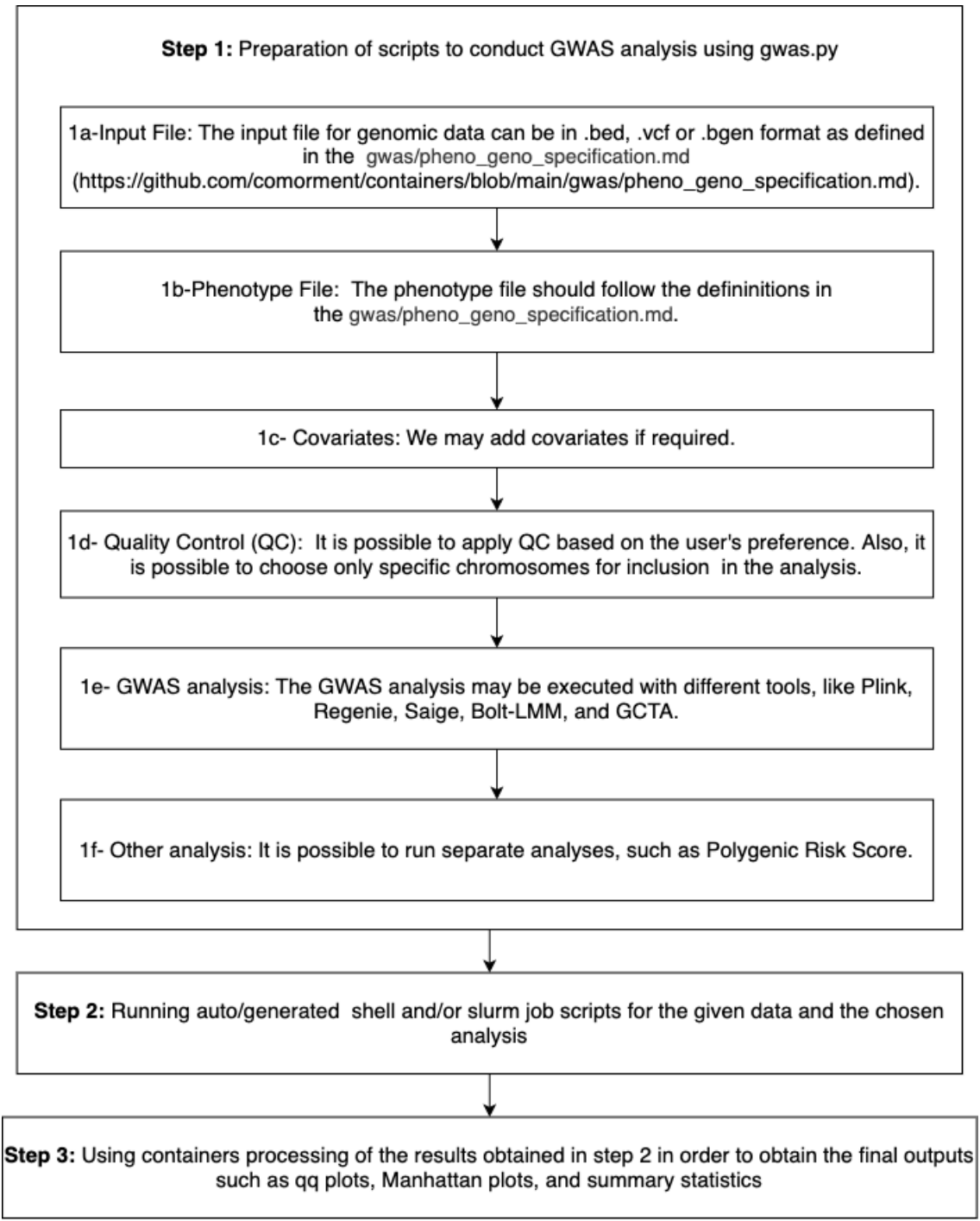

Workflow 1. Running GWAS analysis with COGEDAP

### d) Use case: GWAS analysis using different tools

In this section, we describe how to perform GWAS analysis with genotype and phenotype data formatted according to the specifications presented in the gwas/pheno_geno_specification.md file. As described above, we rely on software containers and the helper gwas.py Python script to read the phenotype data, extract user-defined subsets of phenotypes and covariates, and to prepare the scripts (to run on PC) or SLURM jobs (to run on HPC) for GWAS analysis using the plink2 and regenie software tools. We use example data from reference/examples/regenie folder. In order to simplify scripting, we define environment variables specifying paths, such as "$SIF" to set the path of the container. The path "/REF" is a mounted host directory seen by the running container. These are explained in detail in the Getting started section of the main README file.

For genetic data, we use hard genotype data called in plink format [23], with $n$=500 individuals (reference/examples/regenie/example_3chr.fam) and $m$=500 SNPs across three chromosomes (reference/examples/regenie/example_3chr.bimexample_3chr.bim). In order to run the demo analysis defined here, the helper gwas.py script and config.yaml file from the /gwas directory of the repository should be moved to your working directory. The following commands will create shell scripts or slurm job scripts for a case/control GWAS with plink2 (example 1) and a GWAS for quantitative traits with regenie (example 2). Note that these choices are illustrative; a case/control GWAS could also be run with regenie, and a quantitative trait with plink2 by choosing the appropriate command-line arguments.

Example1: case/control GWAS with plink2

```
export PYTHON="singularity exec --home $PWD:/home $SIF/python3.sif python"

$PYTHON  gwas.py gwas \
--argsfile /REF/examples/regenie/example_3chr.argsfile \
--pheno CASE CASE2 --covar PC1 PC2 BATCH --analysis plink2 figures --out run1_plink2
```

Example2: GWAS for quantitative traits with regenie

```
$PYTHON gwas.py gwas \
--argsfile /REF/examples/regenie/example_3chr.argsfile \
--pheno PHENO PHENO2 --covar PC1 PC2 BATCH --analysis regenie figures --out run2_regenie
```

The gwas.py script takes several command line arguments and values. The mandatory --argsfile argument here points to example_3chr.argsfile which specifies some long flags used across all invocations of the gwas.py script. It defines what phenotype file to use (--pheno-file), which chromosome labels to use (--chr2use), which genotype file to use in fitting the regenie model (--geno-fit-file), as well as the genotype file to use when testing for associations (--geno-file); the --variance-standardize will apply a linear transformation to all continuous phenotypes so that they became zero mean and unit variance, similar to the --variance-standardize argument in plink2. The --info-file points to a file with two columns, SNP and INFO, listing imputation info score for the variants. This is optional and only needed when specifying the the --info threshold argument. Other available QC filters include --maf, --geno and --hwe. Note that in this demo example (example_3chr.argsfile ), --geno-fit-file points to the same file as --geno-file. For real applications, the --geno-fit-file should point to a single genetic file (merged across chromosomes), preferably constrained to less than a million SNPs, for example including only genotyped (rather than imputed) SNPs or constrained to the set of HapMap3 SNPs [11].

Adding the "figures" option to the list of --analysis arguments will trigger post-GWAS scripts to visualize the output of GWAS analysis with manhattan and quantile-quantile (QQ)plots. For this small-scale demo example, the actual GWAS can be executed on a local machine by running auto-generated bash scripts. On an HPC, using gwas/config.yaml to define the specifications of HPC and determine the desired resource allocation, the SLURM job scripts that are generated by gwas.py script and can be submitted in the standard way as given in the file usecases/gwas_demo.md. As a result, these GWAS analyses above will generate GWAS summary statistics files that are merged across all available chromosomes and have a minimal set of columns (SNP, CHR, BP, A1, A2, N, Z, BETA, SE, PVAL).

After all jobs/scripts are executed and summary statistics are obtained, it is possible to plot QQ plots and Manhattan plots using our post-GWAS tools. We also present a real-world application, GWAS in Norwegian Mother, Father and Child Cohort Study (MoBa) [24], on height and major depression phenotypes that can be found in the use case described in usecases/gwas_real.md.

## IV. DISCUSSION

We have developed a comprehensive and user-friendly distributed genomic data analysis platform using software container technologies. Users can conduct GWAS and other analyses on their genome data by using either their local computer or an HPC. Our data analysis platform incorporates a host of different tools, compared to other existing pipelines for genomic analyses such as nf-GWAS [1,2], GWASpi [3], Comprehensive-GWAS [4], and RICOPILI [5]. The number of included tools in the COGEDAP analysis platform is continuously growing thanks to user-provided feedback. The solution provided here allows researchers to use the most recent genomic tools without spending significant time on software installation and investigation of the tools. The software containers provide all relevant software in self-contained virtual machines that are all built using the Ubuntu 20.04 (LTS) Linux-based operating system, and rely only on the external dependencies of Singularity for running the containers, a Shell Command Language (sh) compatible shell (a Linux or Unix-like terminal application such as GNU Bash), and a job scheduler (e.g., SLURM) for submitting jobs to the HPC resource running the actual computing tasks. Some steps have also been made to ensure that the software environment within each container is consistent, that is, Dockerfiles and installation scripts request versions of each software that are explicitly defined during container rebuilds. Source codes and prebuilt containers are change-tracked using Git (https://git-scm.com) with Git LFS (https://git-lfs.github.com/), hosted publicly and freely under an open-source GPL-v3 license on GitHub, ensuring full transparency into their development history. GitHub is presently the main hub for issue tracking, coordinating the development, and running continuous integration (CI) tasks.

In terms of usage, projects, such as the MoBaPsychGen pipeline has already applied the COGEDAP platform for pre-imputation QC, phasing, imputation, and post-imputation [14]. The COGEDAP data analysis platform is also actively being used in several projects, especially from Nordic countries, and has started to gain interest from other countries and projects to conduct genomic analysis. According to the feedback from the researchers using the platform, COGEDAP will keep growing and remain up to date.

Although COGEDAP can be used both on local computers and HPC, as mentioned before, Singularity containers support is a must to run COGEDAP. Therefore, for the systems without Singularity support, it is not possible to run our platform for now. As a future work, we may consider enabling Docker support of COGEDAP. Since we have already developed our containers using Docker, such an extension is quite feasible. Another alternative is adding support for Apptainer (https://apptainer.org/), a fork of Singularity. One possible extension would be to present our platform fully automated as a web-based server so that users may not need to run any script to do their analysis.

Given this interest and current use cases, we believe that our data platform will be an important and helpful tool for scientists to perform various genomic data analyses without spending time on technical details. Having such an automated genomic analysis platform also enables standardization and this would make comparative and/or cooperative analyses easier.

## REFERENCES

[1] Song, Z., Gurinovich, A., Federico, A., Monti, S., & Sebastiani, P. (2021). nf-gwas-pipeline: A Nextflow Genome-Wide Association Study Pipeline. *Journal of Open Source Software*, *6*(59), 2957.

[2] https://genepi.github.io/nf-gwas/about.html

[3] Muñiz-Fernandez, F., Carreño− Torres, A., Morcillo-Suarez, C., & Navarro, A. (2011). Genome-wide association studies pipeline (GWASpi): a desktop application for genome-wide SNP analysis and management. *Bioinformatics*, *27*(13), 1871-1872.

[4] Dagasso, G., Yan, Y., Wang, L., Li, L., Kutcher, R., Zhang, W., & Jin, L. (2020, December). Comprehensive-GWAS: a pipeline for genome-wide association studies utilizing cross-validation to assess the predictivity of genetic variations. In *2020 IEEE International Conference on Bioinformatics and Biomedicine (BIBM)* (pp. 1361-1367). IEEE.

[5] Lam, M., Awasthi, S., Watson, H. J., Goldstein, J., Panagiotaropoulou, G., Trubetskoy, V., ... & Ripke, S. (2020). RICOPILI: rapid imputation for COnsortias PIpeLIne. *Bioinformatics*, *36*(3), 930-933.

[6] Purcell, S., Neale, B., Todd-Brown, K., Thomas, L., Ferreira, M. A., Bender, D., ... & Sham, P. C. (2007). PLINK: a tool set for whole-genome association and population-based linkage analyses. *The American journal of human genetics*, *81*(3), 559-575.

[7] Chang, C. C., Chow, C. C., Tellier, L. C., Vattikuti, S., Purcell, S. M., & Lee, J. J. (2015). Second-generation PLINK: rising to the challenge of larger and richer datasets. *Gigascience*, *4*(1), s13742-015.